\documentclass[12pt]{iopart}
\usepackage{color}

\usepackage{bm}
\usepackage{amsfonts}
\usepackage{graphicx}
\usepackage{caption}
\usepackage[font={small,it}]{caption}
\usepackage{epstopdf}
\usepackage{multirow}
\usepackage[toc,page]{appendix}
\usepackage{rotating}

\usepackage{color}   %May be necessary if you want to color links
\usepackage[linktocpage=true]{hyperref}
\usepackage{hyperref} 
\hypersetup{colorlinks, citecolor=green, filecolor=black, linkcolor=blue, urlcolor=blue }

\expandafter\let\csname equation*\endcsname\relax
\expandafter\let\csname endequation*\endcsname\relax
\usepackage{amsmath}
\numberwithin{equation}{section}

\usepackage{xcolor}
\colorlet{ColorPink}{red!10}
\colorlet{ColorPink}{purple!10}
\colorlet{ColorPink}{magenta!10}
\usepackage{amsfonts, amssymb}
\usepackage{tikz}
\usetikzlibrary{patterns}

\usepackage{fancyhdr}
\usepackage{color}   %May be necessary if you want to color links
\usepackage[linktocpage=true]{hyperref}
\usepackage{hyperref} 
\hypersetup{colorlinks, citecolor=green, filecolor=black, linkcolor=blue, urlcolor=blue }

\expandafter\let\csname equation*\endcsname\relax
\expandafter\let\csname endequation*\endcsname\relax
\usepackage{amsmath}
\numberwithin{equation}{section}
\usepackage{graphicx}

\begin{document}

\title[Finite size corrections to scaling of the formation probabilities and the Casimir effect]
{Finite size corrections to scaling of the formation probabilities and the Casimir effect in the conformal field theories}

\author{M.~A.~Rajabpour}
\address{ Instituto de F\'isica, Universidade Federal Fluminense, Av. Gal. Milton Tavares de Souza s/n, Gragoat\'a, 24210-346, Niter\'oi, RJ, Brazil}

\date{\today{}}

\begin{abstract}
We calculate formation probabilities of the ground state of the finite size quantum critical chains using conformal field theory (CFT) techniques.
In particular, we calculate the formation probability of  one interval in the finite open chain and also formation probability of two disjoint intervals
in a finite periodic system. The presented formulas can be also interpreted as the Casimir energy of needles in particular geometries. We numerically check the validity of
the exact CFT results in the case of the transverse field Ising chain.
\end{abstract}
%\pacs{03.67.Mn,11.25.Hf, 05.70.Jk }
\maketitle
%\tableofcontents

\section{Introduction}

Recently there has been  growing interest in characterizing the formation probabilities in quantum spin chains near quantum critical points
\cite{Stephan2009,Stephan2013,rajabpour2015d,Pozsgay,NR2015,Eisler,Pronko}. The formation probability is defined as
the probability of occurrence of a particular configuration for a given observable in a subsystem. For example, in the case of spin chains, one can think about the probability of having a particular
configuration for a string of spins. A special case of the formation probabilities is called emptiness formation probability. In the spin chains, it is defined as the probability of having
a configuration with all the spins up in the subsystem. Emptiness formation probability of the discrete systems has been studied extensively in the literature, see for example 
\cite{Essler,U1,Abanov-Korepin,Shiroishi,Franchini,Abanov2005}, for the evolution of the emptiness formation probability after a quantum quench see \cite{NR2015,Eisler}.
The main result is that the emptiness formation probability of a simple subsystem of a critical model usually
decreases exponentially with the size of the subsystem. The exponential decay is accompanied by a power-law decay with a universal 
exponent. The nature of this exponent, which is dependent on the central charge of the underlying conformal field theory, has been understood in \cite{Stephan2013} by using 
the boundary conformal field theory (BCFT) techniques.
The main idea is that fixing observables in part of the system is like introducing a slit-like boundary in the euclidean version of the quantum spin chain. If the boundary 
condition induced by the configuration respects
the conformal symmetry, then one can calculate the formation probability using the BCFT techniques. For many other discussions regarding  the BCFT in the presence of a slit geometry see \cite{StephanDubail}.
The analysis is generalized to two-disjoint intervals in \cite{rajabpour2015d} and it was pointed out that by calculating this quantity one can not only derive the central charge of the system
but also, the
whole critical exponents present in the system. It is worth mentioning that in \cite{NR2015}, it was shown numerically that the BCFT results are not only valid for emptiness formation probabilities but also
for much more general formation probabilities. Note that system with the $U(1)$ symmetry  should be studied differently. In this case 
the decay is Gaussian, see for example \cite{Abanov2005,Stephan2013,Allegra2016}. The formation probabilities have also been used to calculate interesting quantities, such as Shannon 
information  \cite{Stephan2009,Shannon information} and mutual Shannon information \cite{mutual information}. 

In apparently unrelated studies, the Casimir effect of two floating objects in a two-dimensional classical medium has been the subject of intense studies for decades, see
for example \cite{Fisher,Gambassi,CasimirCFTolder,Casimir1,Casimir2,Machta,Kardar,needle,Burkhardt2016,Casimir,Dietrich2016,Gambassi2} and references therein. The calculation of the Casimir energy  
of two floating objects in a medium boils down to calculating
the change in the free energy of the medium when there is a special boundary condition in the domain where we positioned the objects. Although normally calculating the Casimir energy is a
difficult task for generic systems it was shown in \cite{Machta,Kardar} that if the system is at the critical point and the boundary conditions are conformal invariant one can 
exactly calculate the Casimir energy of two arbitrary objects using BCFT techniques. 

It is not difficult to see that there is an obvious connection between the emptiness formation probabilities in the quantum spin chains and the Casimir effect of two slits (needles). This connection 
first highlighted in the paper \cite{rajabpour2015d},
in the case of infinite chains. Of course, this connection is not only present in the $1+1$ dimensional systems it can be also seen in arbitrary dimensions.  This simple relation can have numerous mutual benefits for 
these two apparently unrelated areas of research. In \cite{rajabpour2015d}, we have used the formulas presented in the Casimir effect 
studies to calculate the formation probability of two disjoint intervals in an infinite system.
For general formulas regarding Casimir energy of two needles see \cite{Burkhardt2016}.
In this work, we extend the results to finite systems. We study two different cases: 1) formation probability of one interval inside an open finite system and 2) formation probability of two
disjoint intervals in a finite periodic system. The calculations also provide a formula for the Casimir free energy of a needle inside a strip and the Casimir energy of two aligned needles in a cylinder.

The paper is organized as follows: In the next section, we give a brief review of the connection between formation probabilities and the Casimir effect. We also provide all the necessary definitions and formulas.
In the third section, we find a conformal map from strip with a slit to an annulus and then we use it to obtain a formula for the formation probability of a domain inside a 
finite chain. In the fourth section, we study the formation probability
of two disjoint intervals in a finite periodic system. In the fifth section, we numerically check the validity of the presented CFT formulas for the critical transverse field Ising chain.
Finally, in the last section we summarize our findings and we also comment on the future directions.

\section{Formation probabilities and the Casimir effect in CFTs}

In this section, we summarize the main result of the paper \cite{rajabpour2015d} which highlights a relation between the Casimir effect and the formation
probabilities in the quantum chains. The work was based on the earlier studies on the emptiness formation probability of one interval \cite{Stephan2013} and the Casimir effect of arbitrary two dimensional
objects \cite{Machta,Kardar}. Although the relation can be explained most easily in the context of the quantum spin chains the
conclusions are valid for any quantum system in any dimension.
Consider a quantum  system with the Hamiltonian $H$. The transfer matrix of the system is  $T=e^{-\epsilon H}$, where $\epsilon$ is 
the imaginary time step. In the limit of 
infinite steps $N\to\infty$ we have $T^N\sim e^{-\epsilon NE_g}|\psi_g\rangle\langle\psi_g|$. Now imagine we are interested in 
the probability of having an arbitrary configuration $|\mathbf{\sigma}\rangle=|\uparrow\downarrow...\uparrow\rangle_{b}$ in part of the system.
From now on, we call this quantity the formation probability.
Note that the chosen spins do not need to be neighbours. For example, we can think about two disjoint intervals or two disconnected regions
with fixed spins.
One can calculate the formation probability 
by the following formula:
\begin{eqnarray}\label{formation probabilities definition}
 p_{D}=lim_{_{N\to\infty}} \frac{\langle\psi|T^{\frac{N}{2}}\delta(|\mathbf{\sigma}\rangle-|\uparrow\downarrow...\uparrow\rangle_{b})T^{\frac{N}{2}}|\psi\rangle}{\langle\psi|T^N|\psi\rangle},
\end{eqnarray}
where $\delta(|\mathbf{\sigma}\rangle-|\uparrow\downarrow...\uparrow\rangle_{b})$ fixes the spins in the basis $b$ in some arbitrary
directions and $|\psi\rangle$ is the state at infinity
and in principle it can be 
 any state which is not orthogonal to
the ground state. The denominator of the equation (\ref{formation probabilities definition}) 
is  the total partition function of the corresponding $d+1$ dimensional classical system. The infinite system can be subject to periodic (open)
boundary conditions, in that case one needs to work on the infinite cylinder (strip) in two dimensions. In higher dimensions, one needs to think about a generalized cylinder 
or a generalized strip. The numerator is the partition function of the system with this constraint that the spins in the corresponding sub-domain $D$
have particular values. Using the above argument  we can write
\begin{eqnarray}\label{formation probabilities partition function}
 p_{D}=\frac{Z^{D}_b}{Z},
\end{eqnarray}
where $Z^{D}_b$ is the partition function of the whole system minus the domain $D$ and $Z$ is the total partition function. 
Note that the numerator is dependent on the basis that one chooses simply because working on different bases 
induces different boundary conditions on the slit. Instead of  working directly with the formation probabilities,
which usually decays exponentially with respect to the volume,
it is much better to work with its logarithm. We call $\Pi_{D}:=-\ln p_{D}$, the logarithmic formation probability and it is dependent
on the difference between the free energy of the total system and  the free energy of the whole system minus the domain $D$.
It usually follows a volume law with some universal subleading terms that we are interested in calculating them.
When the system is at the critical point these subleading terms can be calculated using  the 
boundary CFT techniques. From now on, we write 
\begin{eqnarray}\label{formation probabilities volume law}
 \Pi_{D}=\alpha V_D+ \Pi_{D}^{CFT}+\gamma,
\end{eqnarray}
where $V$ is the volume of the domain $D$ and $ \Pi_{D}^{CFT}$ is the universal part of the formation probability that can be calculated using CFT techniques  and $\gamma$ is a constant. 
It is not difficult to see that
the $\Pi_{D}^{CFT}$ can be also interpreted as the universal part of the Casimir energy of the system. 
For example, consider an open
chain with a string of fixed observables inside of it. 
In this case, the classical counterpart of the setup is a strip with a needle inside. If we change the slit position, then only  the $ \Pi_{D}^{CFT}$ 
part of the logarithmic formation probability
will change. In other words, the change in  the Casimir free energy  of the needle comes from this part. This is the reason why ultimately 
the Casimir force is just dependent on this term. To see the connection more explicitly it is much better to define the following quantity
\begin{eqnarray}\label{R open}
\mathcal{R}_o:=-\ln\frac{p_{A,B}}{p_A}=-\ln\frac{Z_{A,B}}{Z_A},
\end{eqnarray}
where $A$ and $B$ are the boundary conditions on the slit and the natural boundary of the system, see Figure 1. $Z_{A,B}$ is the partition functions of the strip with a slit inside
and $Z_A$ is the partition function of  an infinite system with a slit inside.
The right-hand side of the equation (\ref{R open}) is the  energy required to bring the boundary from infinity to a finite distance from the needle.

 %%%%%%%%%%%%%%%%%%%%%%%%%%%%%%%%%%%%%%%%%%%%%%%%5%%%%%%%%%%%%%%%%%%%%%%%%%
\begin{figure} [htb] \label{fig0}
\centering
\begin{tikzpicture}[scale=0.6]
\draw [fill=pink!30 ,ultra thick, pink!50] (0,0) rectangle (12,6);
%\draw [fill=magenta!50,  thick, magenta!50] (0.5,5.5) circle [radius=0.4];
%\node [above right, blue] at (0.2,5.25){\large$z$};
\draw [line width=0.1cm ,teal!70] (0,0) -- (12,0);
\draw [line width=0.1cm ,teal!70] (0,6) -- (12,6);
%\draw [darkgray,ultra thick] (6,6) -- (6,4);
\draw [fill=white] (6.10,1.5) rectangle (5.90,4.);
\draw [blue!70,ultra thick] (6.10,1.5) rectangle (5.90,4.);
%\draw [darkgray,ultra thick] (6,1.5) -- (6,0);
%\node [above right, black] at (6.15,4.8){\large$\bar{B}$};
%\node [above right, black] at (6.15,2.5){\large$A$};
%\node [above right, black] at (6.15,0.7){\large$B$};
%\node [above right, black] at (5.25,3.80){\small$P_{2}$};
\node [above right, black] at (6.0,1.25){\tiny\begin{turn}{-90}AAAAAAAA\end{turn}};
\node [above right, black] at (5.3,1.25){\tiny\begin{turn}{90}AAAAAAAA\end{turn}};
%\node [above right, black] at (5.25,1.3){\small$P_{1}$};
\node [above right, black] at (-0.25,-0.6){\tiny$BBBBBBBBBBBBBBBBBBBBBBBBBBBBBBBBBBB$};
\node [above right, black] at (-0.25,5.8){\tiny$BBBBBBBBBBBBBBBBBBBBBBBBBBBBBBBBBBB$};
%\draw [ultra thick,<->] (11.5,0) -- (11.5,6);
%\node [above right, black] at (10.5,3){\large$L$};

\end{tikzpicture}
\caption{(Color online) Strip with a slit inside. The conformal boundary conditions on the two sides of the strip is equal to $B$
and the boundary condition on the slit is $A$.} 
\end{figure}

%%%%%%%%%%%%%%%%%%%%%%%%%%%%%%%%%%%%%%%%%%%%%%%%%%%%%%%%%%%%%%%%%%%

The same argument goes more or less smoothly  also for a periodic system with two slits inside of it. In this case, we define \cite{rajabpour2015d}
\begin{eqnarray}\label{R periodic}
\mathcal{R}_p:=-\ln\frac{p_{A,B}}{p_Ap_B}=-\ln\frac{Z_{A,B}Z}{Z_AZ_B},
\end{eqnarray}
where again $A$ and $B$ are the boundary conditions on the slits and  $Z_{A,B}$, $Z_A$ and $Z_B$ are the partition functions of the cylinder with two slits, slit $A$ and slit $B$ respectively.
The right-hand side of the above equation is the  Casimir free energy of two needles.
In both of the above examples we have two boundaries and, in principle,  they are topologically equivalent to 
 a finite cylinder.
Since we know  the partition function of the
CFT on the cylinder, one 
can hire the techniques known in the boundary CFT to calculate $ \Pi_{D}^{CFT}$. This procedure
has been carried out for generic CFT's in two dimensions in \cite{Machta,Kardar}. The final result
is 
\begin{eqnarray}\label{formation probabilities volume law}
\mathcal{R}_{p(o)}=\mathcal{F}_{ann}+\mathcal{F}_{geo},
\end{eqnarray}
where $\mathcal{F}_{ann}$ is the free energy of the CFT on the annulus which is known for most of the CFT's \cite{Cardy}
and the $\mathcal{F}_{geo}$ can be calculated using standard CFT formulas, see for example \cite{Kardar}:
\begin{eqnarray}\label{integral}
\frac{\delta \mathcal{F}_{geo}}{\delta l}=-\frac{ic}{12\pi}\oint_{\partial S_2}\{w,z\}dz,
\end{eqnarray}
where $\partial S_2$ is a contour surrounding one of the slits, $w$ is the conformal map which maps the original system with two boundaries to an annulus
and $\{w,z\}=\frac{ w'''}{w'}-\frac{3}{2}(\frac{w''}{w'})^2$ is the Schwarzian derivative. In the above equation,
the derivative is taken with respect to the distance between the slits in the problem of two slits in a periodic system which finally gives us the Casimir energy between the two slits, see Figure 2. 
However, logarithmic formation probability
has also another universal term which is a property of every slit without considering the other one. This part of the logarithmic formation probability can be also calculated by using the above formulas, just we need to
consider the derivative appearing in the equation (\ref{integral}) with respect to the length of the slit as it is done also in \cite{Stephan2013}. For the open chain with a slit,
we will consider this case. 
%%%%%%%%%%%%%%%%%%%%%%%%%%%%%%%%%%%%%%%%%%%%%%%%5%%%%%%%%%%%%%%%%%%%%%%%%%
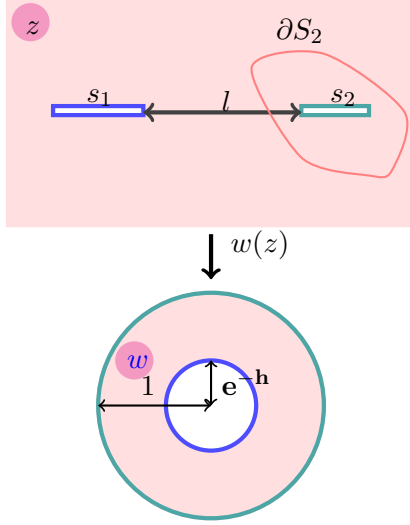
\begin{figure} [htb] 
\centering
\begin{tikzpicture}[scale=0.6]
\draw [fill=white ,ultra thick, pink!50] (0,0) rectangle (9,5);
\draw [fill=magenta!50,  thick, magenta!50] (0.5,4.5) circle [radius=0.4];
%\draw [darkgray,ultra thick] (0,2.5) -- (1,2.5);
\draw [fill=white] (1,2.44) rectangle (3,2.64);
\draw [blue!70,ultra thick] (1,2.44) rectangle (3,2.64);
%\draw [darkgray,ultra thick] (8,2.5) -- (9,2.5);
\draw [fill=white] (6.5,2.44) rectangle (8,2.64);
\draw [teal!70,ultra thick] (6.5,2.44) rectangle (8,2.64);
\draw [darkgray,ultra thick,<->] (3,2.5) -- (6.5,2.5);
\node [above right, black] at (0.15,4.0){\small$z$};
%\node [above right, black] at (0.25,2.5){\large$\bar{B}$};
%\node [above right, black] at (1.7,2.65){\tiny$AAAAAAAAA$};
%\node [above right, black] at (4.3,2.5){\large$B$};
%\node [above right, black] at (6.9,2.65){\tiny$BBBBBBBBB$};
\node [above right, black] at (5.7,3.8){\small$\partial{S_2}$};
\node [above right, black] at (1.5,2.4){\small$s_{1}$};
\node [above right, black] at (4.5,2.3){\small$l$};
\node [above right, black] at (6.9,2.4){\small$s_{2}$};
\draw[thick,red!50] plot [smooth cycle] coordinates {(8.0,1.1)(8.5,1.2)(8.79,2.5)(7,3.8)(5.4,3.5)(6.5,1.5)};

\draw [fill=pink!30 ,ultra thick, pink!50] (4.5,-4) circle [radius=2.5];
\draw [ ultra thick,teal!70 ] (4.5,-4) circle [radius=2.5];
\draw [fill=magenta!50,  thick, magenta!50] (2.8,-3) circle [radius=0.4];
\draw [fill=white ] (4.5,-4) circle [radius=1];
\draw [ultra thick,blue!70] (4.5,-4) circle [radius=1];
\node [above right, blue] at (2.4,-3.4){\small$w$};
\draw [ultra thick,->] (4.5,-0.2) -- (4.5,-1.2);
\node [above right, black] at (4.7,-1.){\small$w(z)$};
\draw [black,thick,<->] (4.5,-4) -- (4.5,-3);
\draw [black,thick,<->] (4.5,-4) -- (2,-4);
\node [above right, black] at (4.5,-4){\small$\bf{e^{-h}}$};
\node [above right, black] at (2.7,-4){\small$1$};

\end{tikzpicture}
\caption{(Color online) Mapping between  a domain with two boundaries (slits) to
an annulus  by the conformal map $w(z)$. The contour $\partial S_2$ is shown by the continuous red line. } 
\end{figure}

The annulus contribution part for a generic CFT with the boundary conditions $A$ and $B$ on the two boundaries
has the following form \cite{Cardy}

\begin{eqnarray}\label{annulus part3}
\ln Z^{annu}(q)&=&\ln [q^{-c/24}(1+\sum_jn^{AB}_jq^{\Delta_j})]-c\frac{h}{12},\\
\label{annulus part4}
\ln Z^{annu}(\tilde{q})&=&\ln [\tilde{q}^{-c/24}(b^{A}_{0}b^B_{0}+\sum_jb^{A}_{j}b^B_{j}\tilde{q}^{\Delta_j})]-c\frac{h}{12},
\end{eqnarray}
where the sum in the first and the second formulas are over the highest weight non-trivial conformal operators and their decendents
which propagate around and across the annulus respectively. We sigled out the contribution of the identity operator in both formulas.
$n^{AB}_j$'s are some non-negative integers and $b^A_{j}=\langle A|j\rangle\rangle$ and $b^B_{j}=\langle\langle j|B\rangle$ with 
$|A(B)\rangle $ and $|j\rangle\rangle$ are Cardy
and Ishibashi states respectively. Different coefficients are related to each other with the formula $n_j^{AB}=\sum_{j'}S_j^{j'}b_j^{A}b_{j'}^B$,
where $S_j^{j'}$ is the element of the modular matrix $S$,  see \cite{Cardy}. Finally,
$q$ and $\tilde{q}$ are defined as
\begin{eqnarray}\label{q tilde q}
q=e^{-\pi\frac{2\pi}{h}},\hspace{1cm}\tilde{q}=e^{-2h},
\end{eqnarray}
where $e^{-h}$ is the inner radius of the annulus and the outer radius is $1$. As it is clear, for large $h$ the expansion with respect to 
$\tilde{q}$, i.e. (\ref{annulus part4}) is useful and when $h$ is small the expansion with respect to $q$, i.e. (\ref{annulus part3}).

An equivalent way of writing the equations (\ref{annulus part3}) and (\ref{annulus part4}) is
\begin{eqnarray}\label{annulus CFT}
\mathcal{F}_{ann}=-\ln Z_{AB}(\tilde{q})=c\frac{h}{12}-\ln \sum_{\Delta}b_{\Delta}^Ab_{\Delta}^B\chi_{\Delta}(\tilde{q}),\\
\mathcal{F}_{ann}=-\ln Z_{AB}(q)=c\frac{h}{12}-\ln \sum_{\Delta}n_{\Delta}^{AB}\chi_{\Delta}(q),
\end{eqnarray}
where $\chi_{\Delta}(\tilde{q})(\chi_{\Delta}(q))$ is the character of the conformal operator with the conformal weight $\Delta$.

\section{One interval in an open system}

In this section, we study the logarithmic formation probability of a string with the length $l$ inside a finite chain with the total length $L$. 
In two dimensions the strip with the slit can be mapped to an annulus and later to a cylinder. In principle the boundary 
conditions on the slit and on the natural boundaries of the open system can be different which means that the two boundaries of the cylinder
can have different conditions. This subtlety can be easily considered in the framework of our previous section.

The  setup that we would like to study is
shown in the Figure ~3.  To derive the logarithmic formation probability, one needs to calculate the partition
function of the  surface shown in the the Figure ~3. The corresponding conformal map from strip with one slit  to a cylinder can be derived as follows:
%\begin{figure} [htb] \label{fig1}
%\center
%\includegraphics[width=1.2\textwidth]{OBC.eps}
%\caption{(Color online) Mapping between different regions. The strip with the slit  can be mapped to
%cylinder  by the conformal map $\tilde{w}$ in four steps. } 
%\end{figure}
 %%%%%%%%%%%%%%%%%%%%%%%%%%%%%%%%%%%%%%%%%%%%%%%%5%%%%%%%%%%%%%%%%%%%%%%%%%
\begin{figure} [htb] \label{fig2}
\centering
\begin{tikzpicture}[scale=1]
\draw [fill=pink!30 ,ultra thick, pink!50] (0,0) rectangle (12,6);
\draw [fill=magenta!50,  thick, magenta!50] (0.5,5.5) circle [radius=0.4];
\node [above right, blue] at (0.2,5.25){\large$z$};
\draw [line width=0.1cm ,teal!70] (0,0) -- (12,0);
\draw [line width=0.1cm ,teal!70] (0,6) -- (12,6);
\draw [darkgray,ultra thick] (6,6) -- (6,4);
\draw [fill=white] (6.10,1.5) rectangle (5.90,4.);
\draw [blue!70,ultra thick] (6.10,1.5) rectangle (5.90,4.);
\draw [darkgray,ultra thick] (6,1.5) -- (6,0);
%\node [above right, black] at (6.15,4.8){\large$\bar{B}$};
%\node [above right, black] at (6.15,2.5){\large$A$};
%\node [above right, black] at (6.15,0.7){\large$B$};
\node [above right, black] at (5.25,3.80){\small$P_{2}$};
\node [above right, black] at (6.25,2.60){\large$s$};
\node [above right, black] at (5.25,1.3){\small$P_{1}$};
\node [above right, black] at (6.3,0.6){\large$l$};
\draw [ultra thick,<->] (11.5,0) -- (11.5,6);
\node [above right, black] at (10.5,3){\large$L$};

\draw [ultra thick,->] (3.5,-0.5) -- (3.5,-1.4);
\node [above right, black] at (2.8,-1.1){$I$};

\draw [fill=pink!30 ,ultra thick, pink!50] (3.7,-4.2) circle [radius=2.2];
\draw [ ultra thick,teal!70 ] (3.7,-4.2) circle [radius=2.2];
\draw [fill=white] (2.1,-4.2) rectangle (4.4,-4);
\draw [blue!70,ultra thick] (2.1,-4.2) rectangle (4.4,-4);
\draw [ultra thick,->] (3.7,-4.) -- (3.7,-2.5);
%\draw [ultra thick,dashed,->] (4.4,-4.1) -- (5.9,-4.1);
\node [above right, black] at (1.9,-4.8){\small$P_{2}$};
\node [above right, black] at (4.,-4.8){\small$P_{1}$};
\draw [fill=magenta!50,  thick, magenta!50] (2.5,-3.0) circle [radius=0.3];
\node [above right, blue] at (2.2,-3.3){\large$z_1$};

\draw [ultra thick,->] (6.1,-4.1) -- (7.,-4.1);
\node [above right, black] at (6.2,-3.9){\large$II$};

\draw [fill=pink!30 ,ultra thick, pink!50] (9.5,-4.2) circle [radius=2.2];
\draw [ ultra thick,teal!70 ] (9.5,-4.2) circle [radius=2.2];
\draw [fill=white] (8.3,-4.2) rectangle (10.6,-4);
\draw [blue!70,ultra thick] (8.3,-4.2) rectangle (10.6,-4);
\draw [ultra thick,->] (9.5,-4.) -- (9.5,-2.5);
\draw [ultra thick,dashed,->] (10.6,-4.1) -- (11.7,-4.1);
\node [above right, black] at (8,-4.8){\small$P_{1}$};
\node [above right, black] at (10.2,-4.8){\small$P_{2}$};
\node [above right, black] at (8.8,-4.){\small$d$};
\node [above right, black] at (9.8,-4){\small$d$};
\draw [fill=magenta!50,  thick, magenta!50] (8.3,-3.0) circle [radius=0.3];
\node [above right, blue] at (8,-3.3){\large$z_2$};

\draw [ultra thick,->] (9.5,-6.8) -- (9.5,-7.7);
\node [above right, black] at (8.5,-7.5){\large$III$};

\draw [fill=pink!30 ,ultra thick, pink!50] (9.5,-10.2) circle [radius=2.2];
\draw [ ultra thick,teal!70 ] (9.5,-10.2) circle [radius=2.2];
\draw [fill=white ] (9.5,-10.2) circle [radius=1];
\draw [ultra thick,blue!70 ] (9.5,-10.2) circle [radius=1];
\draw [ultra thick,<->] (9.5,-10.2) -- (9.5,-9.2);
\draw [ultra thick,<->] (9.5,-10.2) -- (7.3,-10.2);
%\draw [ultra thick,dashed] (10.5,-10.2) -- (11.7,-10.2);
\node [above right, black] at (9.5,-10.){\small$\bf{e^{-h}}$};
\node [above right, black] at (7.8,-10.){\large$\bf{1}$};
\draw [fill=magenta!50,  thick, magenta!50] (8.3,-8.8) circle [radius=0.3];
\node [above right, blue] at (8,-9.1){\large$w$};

\draw [ultra thick,->] (6.8,-10.2) -- (5.7,-10.2);
\node [above right, black] at (6.0,-10){\large$IV$};

\draw [fill=pink!30 ,ultra thick, pink!50] (2,-13) rectangle (5,-8);
\draw [fill=magenta!50,  thick, magenta!50] (2.5,-8.5) circle [radius=0.3];
\node [above right, blue] at (2.2,-8.75){\large$\bar{w}$};
\draw [line width=0.1cm ,teal!70] (2,-13) -- (2,-8);
\draw [line width=0.1cm ,blue!70] (5,-13) -- (5,-8);
\node [above right, black] at (1.2,-10.55){\large$2\pi$};
\node [above right, black] at (3.2,-8){\large$h$};
\draw [ultra thick,<->] (2,-8) -- (5,-8);

\end{tikzpicture}
\caption{(Color online) Mapping between different regions. The strip with width $L$ and a slit inside can be mapped to
a cylinder with the length $h$ and circumference $2\pi$ by the conformal map $\tilde{w}(z)$ in four steps.} 
\end{figure}
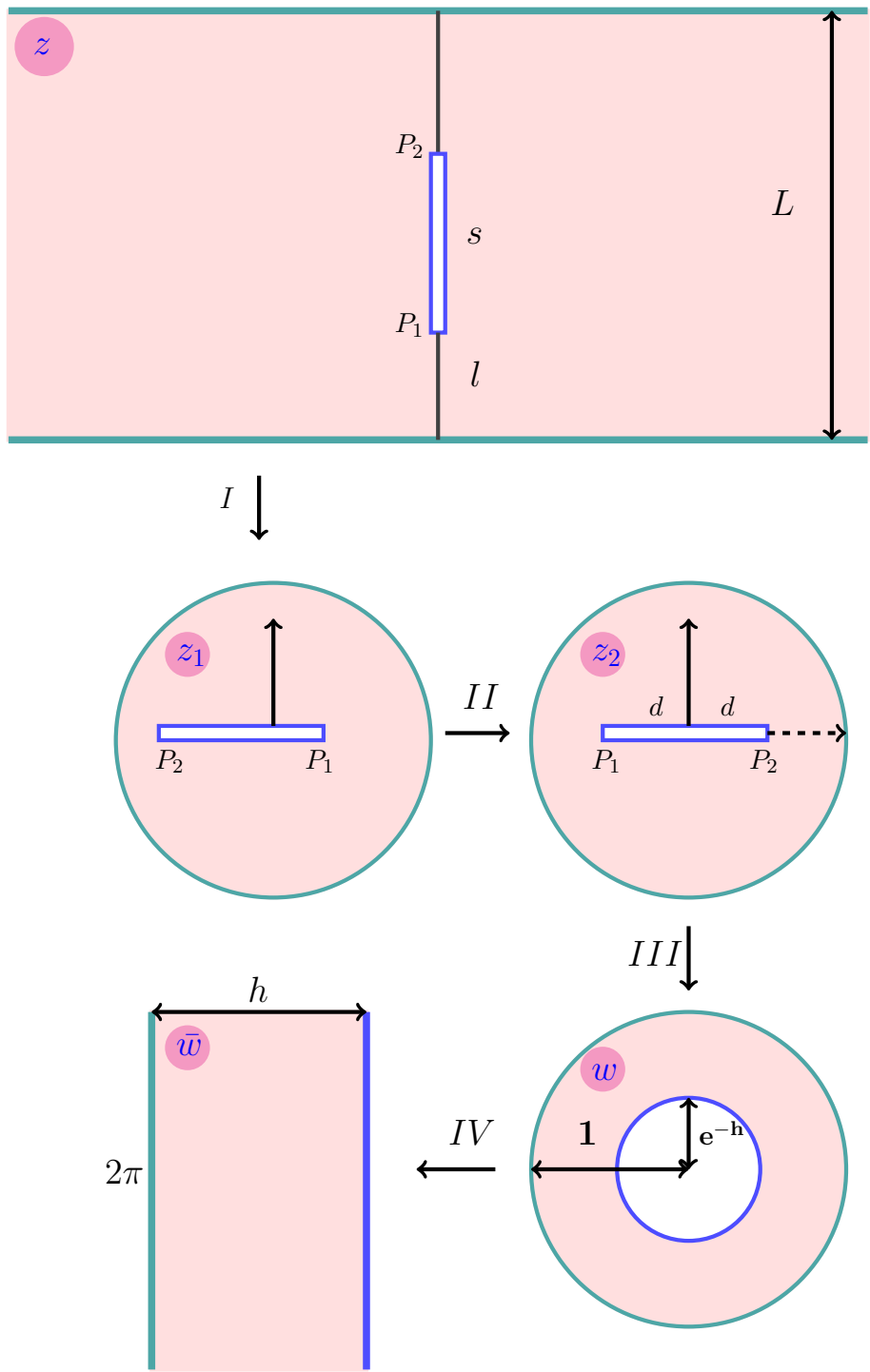

%%%%%%%%%%%%%%%%%%%%%%%%%%%%%%%%%%%%%%%%%%%%%%%%%%%%%%%%%%%%%%%%%%%

Step I: We first map the strip to a unit disc by the conformal map 
\begin{equation}\label{circle to strip}
z_1=i \frac{e^{\frac{\pi z}{L}}-i}{e^{\frac{\pi z}{L}}+i}.
\end{equation}
The coordinates of $P_1$  and  $P_2$ are now $(b,0)$ and $(a,0)$ respectively with
\begin{eqnarray}\label{a and b}
a=\frac{\cos[\frac{\pi l}{L}]}{1+\sin[\frac{\pi l}{L}]},\\
b=\frac{\cos[\frac{\pi (l+s)}{L}]}{1+\sin[\frac{\pi (l+s)}{L}]}.
\end{eqnarray}

Step II: The unit disc with the unsymmetrical slit can be mapped to a unit disc with a symmetrical slit by
the conformal map
\begin{eqnarray}\label{unit to unit}
z_2=\frac{g-z_1}{1-gz_1},\\
g=\frac{1+ab-\sqrt{(a^2-1)(b^2-1)}}{a+b}.
\end{eqnarray}
The length of the slit is now $2d$ with
\begin{equation}\label{d}
d=\frac{-1+ab+\sqrt{(a^2-1)(b^2-1)}}{b-a}.
\end{equation}

Step III : The remaining disc with a slit can now be mapped to an annulus by using the conformal map
$w(z_2)$ provided in \cite{Nehari} as
\begin{equation}\label{conformal map OBC}
w(z_{z_2})=i e^{-h}e^{\frac{\pi}{2i\mathcal{K}(k^2)}\mbox{sn}^{-1}(\frac{z_2}{d},k^2)};
\end{equation}
where $\mathcal{K}$ and $\mbox{sn}^{-1}$ are
the elliptic  and inverse Jacobi functions \footnote{Note that in all of the formulas we adopt
the Mathematica convention for all the elliptic functions.} respectively and
\begin{eqnarray}\label{k and h OBC}
k&=&d^2,\\
h&=&\frac{\pi}{4}\frac{\mathcal{K}(1-k^2)}{\mathcal{K}(k^2)}.
\end{eqnarray}
Note that the equation (\ref{conformal map OBC}) is valid just for $\mbox{Im}z>0$, for the lower half $\mbox{Im}z<0$,
we need to use
\begin{equation}\label{conformal map OBC2}
w(z)=-i e^{-h}e^{\frac{\pi}{2i\mathcal{K}(k^2)}\mbox{sn}^{-1}(-\frac{z_2}{d},k^2)};
\end{equation}

Step IV : The final step is mapping the annulus to a cylinder by $\tilde{w}=\ln w$. Since we already wrote the explicit form of
the partition function of the annulus,  in principle, this step is  redundant. However, it is worth mentioning that the term 
$c\frac{l}{12}$ in the partition function of the annulus is the result of this step.

To calculate the Schwarzian derivative we need the following  chain rule 
\begin{eqnarray}\label{chain rule}
S(f_1\circ f_2)=\big{(}S(f_1)\circ f_2\big{)}(f_2')^2+S(f_2).
\end{eqnarray}
The calculations can be done using Mathematica. Finally, we have
\begin{eqnarray}\label{geometric OBC}
\frac{\delta \mathcal{F}_{geo}}{\delta s}=\frac{\pi^3 c}{192Ld}\frac{1-4gd+g^2+d^2(1+g^2)}{(d^4-1)(g^2-1)\mathcal{K}^2(k^2)}\nonumber\\
+\frac{\pi c}{96Ld}\frac{1+8gd+8gd^5+g^2-5d^2(1+g^2)-5d^4(1+g^2)+d^6(1+g^2)}{(d^4-1)(g^2-1)}.
\end{eqnarray}
For symmetric slit with $l=\frac{L-s}{2}$ we have $b=-a=-d$ and $g=0$. Then the above formula has much simpler form
\begin{eqnarray}\label{geometric OBC symmetric}
\frac{\delta \mathcal{F}_{geom}}{\delta s}=\frac{\pi c}{192Ld}\frac{\pi^2+2(d^4-6d^2+1)\mathcal{K}^2(k^2)}{(1-d^2)\mathcal{K}^2(k^2)}.
\end{eqnarray}
The above equations combined with the $\mathcal{F}_{annu}$ provide us all the necessary formulas for the formation probability of
a slit inside a finite chain. Since the final integral over $s$ can not be calculated analytically, one can rely on numerical calculations.
It is illuminating to study the above formulas in the limit of $L\to\infty$. In this limit, we have
\begin{eqnarray}\label{d}
d_{L\to\infty}\approx \frac{\pi s}{4L},\hspace{1cm}h_{L\to\infty}\approx\ln\frac{8L}{\pi s}.
\end{eqnarray}
In particular, in this limit $\tilde{q}$ is the small parameter with the following asymptotic limit
\begin{eqnarray}\label{tilde q}
\tilde{q}\approx (\frac{\pi s}{8L})^{2}.
\end{eqnarray}
Then  after a bit of calculations we have
\begin{eqnarray}\label{Z geometric L infty}
 \mathcal{F}_{geom}=\frac{c}{8}\ln \frac{s}{a} +...,\\
 \ln Z^{annu}(\tilde{q})=\ln b^{A}_{0}+\ln b^B_{0}+\frac{b^{A}_{1}b^B_{1}}{b^{A}_{0}b^B_{0}}(\frac{\pi s}{8L})^{2\Delta_1},
\end{eqnarray}
where $\Delta_1$ is the smallest  scaling dimension present in the spectrum of the system and $a$
is the UV cutoff. Putting all the terms together we finally have
\begin{eqnarray}\label{formation probabilities OBC large system}
\Pi_{D}^{CFT}= \frac{c}{8}\ln \frac{s}{a}  -\ln b^{A}_{0}-\ln b^B_{0}-\frac{b^{A}_{1}b^B_{1}}{b^{A}_{0}b^B_{0}}(\frac{\pi s}{8L})^{2\Delta_1},
\end{eqnarray}
The first term which is dependent only to the central charge of the system
is already derived in \cite{Stephan2013}. The  constant terms are the Affleck-Ludwig
boundary entropies and the last term is the finite size correction to the formation probability. Note that there are also some further corrections 
coming from the geometric part of the partition function, see \cite{Stephan2013}, however, they are not dependent on the spectrum of the system.
To determine the boundary entropy term one should first remove a piece of non-universal
constant term. This can be done by considering a periodic or infinite system, then we have
\begin{eqnarray}\label{formation probabilities PBC large system}
\Pi_{D}^{CFT}= \frac{c}{8}\ln \frac{s}{a}  -\ln b^{A}_{0}+....
\end{eqnarray}
where again $\ln b^{A}_{0}$ is the boundary contribution of the slit. The boundary contribution of the natural boundary can be derived by subtracting
the two equations (\ref{formation probabilities OBC large system}) and (\ref{formation probabilities PBC large system}). 

\section{Two disjoint intervals in a periodic system}

In this section, we study the logarithmic formation probability in the finite periodic system. The most general case that we are interested in is the one with two disjoint intervals.
When we have one interval the formation probability has been already calculated in \cite{Stephan2013}:
\begin{eqnarray}\label{FP one interval}
\Pi=\frac{c}{8}\ln \Big{(}\frac{L}{\pi}\sin\frac{\pi s}{L}\Big{)}+....
\end{eqnarray}
In the presence of two intervals we have
\begin{eqnarray}\label{FP one interval}
\Pi(L,s_1,s_2,l)=\frac{c}{8}\ln \Big{(}\frac{L}{\pi}\sin\frac{\pi s_1}{L}\Big{)}+\frac{c}{8}\ln \Big{(}\frac{L}{\pi}\sin\frac{\pi s_2}{L}\Big{)}+\mathcal{R}_p(L,s_1,s_2,l),
\end{eqnarray}
\begin{figure} [htb] \label{fig4}
\center
\includegraphics[width=0.6\textwidth]{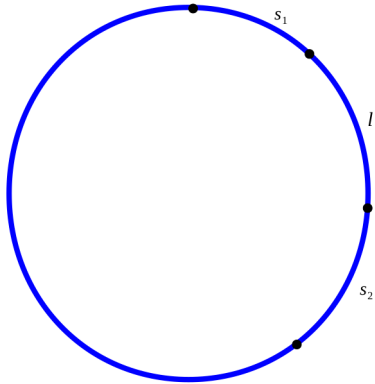}
\caption{(Color online) Setup for two disjoint intervals with lengths $s_1$ and $s_2$ for a system with the total size $L$. } 
\end{figure}
where $s_1$ and $s_2$ are the length of the slits and $\mathcal{R}_p(L,s_1,s_2,l)$ is the part which is dependent on the full spectrum of the system and it is also dependent on 
the  geometrical situations of the two slits, see Figure ~4. Note that $\mathcal{R}_p(L,s_1,s_2,l)$  is
the same as the corresponding Casimir energy. We would like to calculate $\mathcal{R}_p(L,s_1,s_2,l)$ in the presence of two slits \cite{rajabpour2015d} positioned at distance $l$ from each other.
In other words, we are interested in study the Casimir energy of two needles in a periodic system.
The conformal map of a cylinder with two slits to an annulus has been already presented in \cite{Rajabpour2016} and we wrote it explicitly in the appendix.
The Schwarzian derivative and  the integral  over the poles at $z=s_1+l$ and $z=s_1+l+s_2$ can be calculated using the Mathematica.
 Then the final result has  the following form:

\begin{eqnarray}\label{finite size geometric}
\frac{\delta \ln Z^{geom}_{\alpha}}{\delta l}=-i\pi c \frac{P-Q\mathcal{K}^2(1-k^2)}{ R \mathcal{K}^2(1-k^2)},
\end{eqnarray}
with 

\begin{eqnarray}
P=2\pi^2\Big{(}-4k(e^{2\pi i\frac{l+s_1}{L}}-1)+(1+k)^2e^{2\pi i\frac{s_1}{L}}(e^{2\pi i\frac{l}{L}}-1)^2\Big{)},\nonumber\\
Q=(1+6k+k^2)\times\nonumber\\
\Big{(}-2(k-1)^2e^{2\pi i\frac{l+s_1}{L}}-4k-4ke^{4\pi i\frac{l+s_1}{L}}+(1+k)^2e^{2\pi i\frac{s_1}{L}}+(1+k)^2e^{2\pi i\frac{2l+s_1}{L}}\Big{)},\nonumber\\
R=48Lk(1+k)^2(-1+e^{\frac{2i\pi l}{L}})(-1+e^{\frac{2i\pi s_1}{L}})(-1+e^{\frac{2i\pi (s_1+l)}{L}}).\nonumber
\end{eqnarray}
and 
\begin{eqnarray}\label{h general}
h=2\pi\frac{\mathcal{K}(k^2)}{\mathcal{K}(1-k^2)},
\end{eqnarray}
Having the above formulas, now we apply them to three interesting regimes: Two small slits far from each other, one big and one small slits far from each other and two big slits close to each other.

\subsection{ $s_1,s_2\ll l$:} 

When two small slits are far from each other the geometric part of the free energy decays like $\frac{1}{l^5}$ with respect to the distance $l$, see \cite{Kardar}. Because of the fast decay
of this term, usually, we can ignore it in favor of more dominant terms coming from the annulus part. This is true also when one of the slits is big but the other one is small. Since in this case we have
\begin{eqnarray}\label{h PBC 1}
h=-2\ln \frac{\pi s_1s_2}{4L\sin\frac{\pi l}{L}},
\end{eqnarray}
then the small parameter is $\tilde{q}$ with
\begin{eqnarray}\label{tilde q PBC 1}
\tilde{q}=(\frac{\pi s_1s_2}{4L\sin\frac{\pi l}{L}})^4.
\end{eqnarray}
Finally we have
\begin{eqnarray}\label{formation probabilities PBC two small slits}
\mathcal{R}(L,s_1,s_2,l)= -\ln b^{A}_{0}-\ln b^B_{0}-\frac{b^{A}_{1}b^B_{1}}{b^{A}_{0}b^B_{0}}(\frac{\pi s_1s_2}{4L\sin\frac{\pi l}{L}})^{4\Delta_1}.
\end{eqnarray}
Note that if the smallest scaling dimension present in the system is bigger than ~1, we need to consider the geometric part of the free energy as the leading decaying term.

\subsection{ $s_2 \ll s_1, l$ :}   

In this case, we have one big and one small slit. As we discussed in the previous subsection the geometric part can be ignored once again in favor of the annulus part.
The small parameter again is the $\tilde{q}$ and we have
\begin{eqnarray}\label{h PBC 2}
h=-\ln \frac{\pi s_2\sin\frac{\pi s_1}{L}}{16L\sin\frac{\pi l}{L}\sin\frac{\pi(l+s_1)}{L}},\\
\tilde{q}=\Big{(}\frac{\pi s_2\sin\frac{\pi s_1}{L}}{16L\sin\frac{\pi l}{L}\sin\frac{\pi(l+s_1)}{L}}\Big{)}^2.
\end{eqnarray}
Then we can simply write
\begin{eqnarray}\label{formation probabilities PBC one big one small slits}
\mathcal{R}(L,s_1,s_2,l)= -\ln b^{A}_{0}-\ln b^B_{0}-\Big{(}\frac{\pi s_2\sin\frac{\pi s_1}{L}}{16L\sin\frac{\pi l}{L}\sin\frac{\pi(l+s_1)}{L}}\Big{)}^{2\Delta_1}.
\end{eqnarray}
The above formula in the limit of $s_1$ small gives back the equation (\ref{formation probabilities PBC two small slits}).

\subsection{ $l \ll s_1,s_2$ :} 

Here, we have two big slits that are in a small distance from each other. 
First, it is better to send the size of the system to infinity, i.e. $L\to\infty$ and then study the Casimir energy. For simplicity, we also
consider $s_1=s_2=s$. In this case, we have
\begin{eqnarray}\label{geometric s1 s2 equal infinite size}
\frac{\delta \ln Z^{geom}}{\delta l}=\frac{c}{12}\frac{\pi^2(l+2s)^2-2(2l^2+4ls+s^2)\mathcal{K}^2(\frac{4s(l+s)}{(l+2s)^2})}
{l(l+s)(l+2s)\mathcal{K}^2(\frac{4s(l+s)}{(l+2s)^2})}.
\end{eqnarray}
When $l$ is much smaller than $s$ the small parameter is $q$, in other words, we have
\begin{eqnarray}\label{ q}
h=\frac{\pi^2}{\ln\frac{8s}{l}}+...,\hspace{0.6cm}q=(\frac{l}{8s})^{2}+....
\end{eqnarray}
The above equations already appeared in \cite{rajabpour2015d}. Then one can get 
\begin{eqnarray}\label{expansion of geometric part 2 geom}
\ln Z^{geom}=\frac{c}{12}\big{(}\frac{\ln\frac{a}{l}}{2}+\frac{\pi^2}{\ln\frac{8s}{l}}\Big{)}+...,\\
\label{expansion of geometric part 2 annu}
\ln Z^{annu}=\frac{c}{12}\big{(}\ln\frac{8s}{l}-\frac{\pi^2}{\ln\frac{8s}{l}}\Big{)}+n_1(\frac{l}{8s})^{2\Delta_1}+....
\end{eqnarray}
Finally for fixed but large $s$ we have

\begin{eqnarray}\label{fixed large s}
\mathcal{R}^{CFT}(l)=\frac{c}{8}\ln \frac{l}{a}+...,
\end{eqnarray}
where the dots are the the subleading terms including the $s$ dependent terms.

\section{Transverse field Ising chain: numerical calculations }

In this section, we numerically check the validity of some of our analytical formulas for the transverse field Ising chain. The Hamiltonian of the critical Ising chain is  as follows:
\begin{eqnarray}\label{Hamiltonian Ising}
H=-\sum_{j=1}^L\Big{[}\sigma_j^x\sigma_{j+1}^x+\sigma_j^z\Big{]}.
\end{eqnarray}
 Using the Jordan-Wigner transformation the above Hamiltonian can be mapped into a
Hamiltonian of the free fermions. The central charge of the transverse field Ising chain is $c=\frac{1}{2}$.
The numerical procedure to calculate the formation probabilities in the $\sigma^z$ basis is already explained in \cite{NR2015,rajabpour2015d} and we do not report it here. However,
it is worth mentioning that one can calculate the formation probabilities easily by having the correlation functions of the fermionic operators.
The prototypical configurations that respect the conformal symmetry in the Ising chain are the ones with all the spins in the  $\sigma^z$ basis up and the one with 
all the spins in the $\sigma^z$ basis down.  All the spins up 
leads to free boundary conditions with $b_0=1$ and $\Delta_1=\frac{1}{2}$. The argument goes as follows:
The  spins in the two-dimensional classical Ising model are in the  $\sigma^x$ basis, so to find the connection to the 
familiar boundary conditions, we write the configuration of all the $\sigma^z$ spins up in the $\sigma^x$ basis as follows:
\begin{eqnarray}\label{Free BC}
|\uparrow...\uparrow>_{z}=\frac{1}{2^{s/2}}(|\rightarrow>_x+|\rightarrow>_x)...(|\rightarrow>_x+|\rightarrow>_x)\hspace{3cm}\nonumber\\
=\frac{1}{2^{s/2}}\sum_{\{\sigma_j^x\}}|\sigma_1^x...\sigma_s^x>=|\text{free}>.
\end{eqnarray}
The above equation means that, all the spins up configuration should flow to free BCFT.
The formation probability in this special case has a very simple form, it can be written as \cite{Franchini}:
\begin{eqnarray}\label{emptiness formation probability}
p_{\uparrow}=\det[\frac{1-G}{2}],
\end{eqnarray}
where $G$ is the correlation matrix of the free fermions which has the following forms:
  \begin{eqnarray}\label{G Ising PBC }
G^P_{ij}&=&-\frac{1}{L\sin(\frac{\pi(i-j+1/2)}{L})},\\
\label{G Ising OBC }
G^O_{ij}&=&-\frac{1}{2L+1}\Big{(}\frac{1}{\sin(\frac{\pi(i-j+1/2)}{2L+1})}+\frac{1}{\sin(\frac{\pi(i+j+1/2)}{2L+1})}\Big{)}.
\end{eqnarray}
where $G^P_{ij}$ and $G^O_{ij}$ are the correlation matrices for the periodic and open boundary conditions respectively. The above correlation matrix for the open boundary condition is derived by 
considering free boundary conditions. The numerical results indicate that
all the spins $\sigma^z$ down configuration
leads to the fixed boundary condition, see \cite{NR2015}\footnote{ Note that  in \cite{rajabpour2015d} it was shown
that the emptiness formation probability of two disjoint intervals with all the spins down configuration can be derived
from the partition function of the Ising CFT with free boundary conditions. The reason  is that 
although this configuration flow to a fixed boundary condition it is not  clear that the fixed configurations on the two slits are alike or different. This freedom forces us to work
with $Z_{_{Fi1-Fi1}}+Z_{_{Fi1-Fi2}}$ which is the same as $Z_{_{Fr-Fr}}$.}. This fact, will be further supported with the subsequent numerical calculations. The formation probability in this special case has a very simple form, it can be written as \cite{Franchini}:
\begin{eqnarray}\label{emptiness formation probability}
p_{\downarrow}=\det[\frac{1+G}{2}].
\end{eqnarray}
Although there are many other configurations that flow to
 the conformal boundary conditions, here we will only focus on  the  configurations mentioned above. For a discussion regarding all the possible conformal configurations and
 their conformal boundary counterparts see \cite{NR2015,NR2016}. The corresponding annulus partition functions of different boundary conditions are listed in Appendix B.

 First, we have calculated the logarithmic formation probability of an  interval positioned symmetrically inside a finite open system. 
 In the Figures ~5 and ~6, we plot the logarithmic formation probability
 of a slit inside a chain of size $L=300$ minus the logarithmic formation probability of the same slit in an infinite system for the two configurations, all spins up and all spins down. 
 The numerical results are in  agreement with the CFT
 predictions. There is a bit of deviation from the CFT results for large $s$ in the case of all spins down configuration which might be due to extra finite size terms that are usually present
 for this configuration \cite{Franchini}. Note that in the case of all the spins up and down configurations we used $Z_{_{Fr-Fr}}$ and $Z_{_{Fr-Fi}}$ respectively. Their exact forms have been written in the appendix. It is worth mentioning that in both of the above
 cases 
  from the numerical calculations it is clear that the boundary entropy of the natural boundary of the system is equal to zero.
 
 \begin{figure} [htb] \label{fig5}
\center
\includegraphics[width=0.5\textwidth]{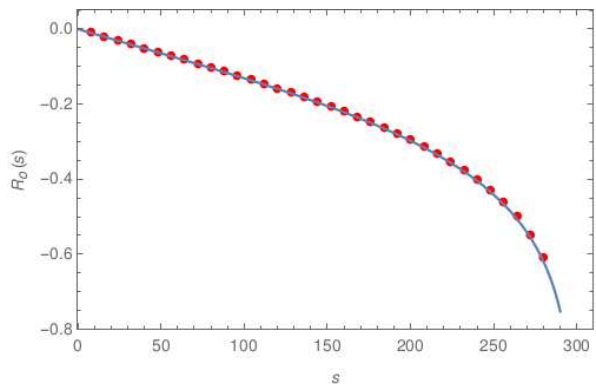}
\caption{(Color online) $\mathcal{R}_o(s)$ with respect to $s$ for the transverse field Ising chain with open boundary conditions for the all spins up configuration. The size of the total system is $L=300$. } 
\end{figure}

 \begin{figure} [htb] \label{fig6}
\center
\includegraphics[width=0.5\textwidth]{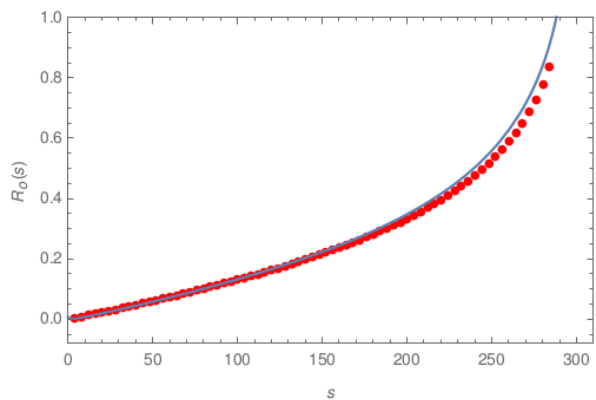}
\caption{(Color online) $\mathcal{R}_o(s)$ with respect to $s$ for the transverse field Ising chain with open boundary conditions for all the spins down configuration. The size of the total system is $L=300$. } 
\end{figure}

We also calculated $\mathcal{R}_p(L,s_1,s_2,l)$ with $s_1=s_2=s$ for the periodic transverse field Ising chain in the presence of all the spins up configuration. 
The results depicted in the Figure ~7 are also 
consistent with the CFT formula provided in the last section.
Finally, we also verified the validity of the equation (\ref{fixed large s}) for large $s$ in the Figure ~8.
 
 \begin{figure} [htb] \label{fig7}
\center
\includegraphics[width=0.5\textwidth]{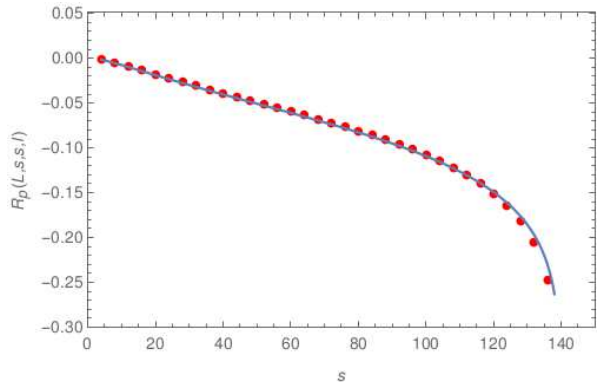}
\caption{(Color online) $\mathcal{R}_p(L,s,s,l)$ with respect to $s$ for the transverse field Ising chain with periodic boundary conditions. The size of the total system is $L=300$ and we used fix $l=20$. } 
\end{figure}

 \begin{figure} [htb] \label{fig8}
\center
\includegraphics[width=0.5\textwidth]{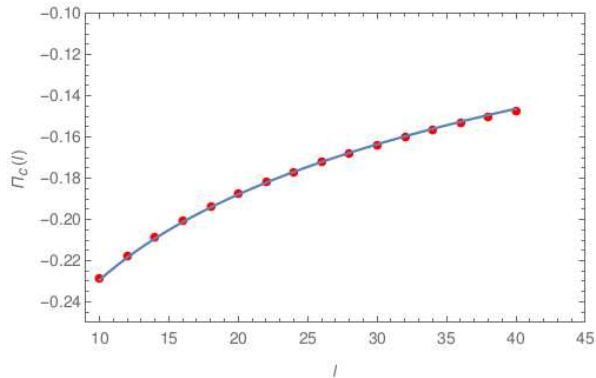}
\caption{(Color online) $\mathcal{R}(l)$ with respect to $l$ for the infinite transverse field Ising chain. We took $s=800$ and the full line is the function $a\ln l+b$ with $a=0.06$ and $b=-0.36$. } 
\end{figure}

\section{Conclusions}

In this paper, we studied the effect of the finite size on the formation probability of the quantum critical chains described by conformal field theories.
First, we have calculated the formation probability of an interval inside an open critical system. Then, we have calculated the formation probability of
two disjoint intervals in a finite periodic system. The studied examples  can be also interpreted as the Casimir energy of a needle inside a strip and the Casimir
energy of two aligned needles on the cylinder. Then, we have checked the validity of the CFT results for the case of the
critical transverse field Ising chain. The numerical results are in perfect agreement with the exact CFT calculations. Since the quantities calculated in this paper are dependent on the 
full structure of the underlying CFT of the system they can be used to fix the universality class of the model.
There are at least three interesting directions to generalize the ideas of this paper.
First of all, it will be interesting to  generalize the analytical results coming from BCFT
 to a critical system in higher dimensions. Secondly, it will be nice to calculate the emptiness formation probability of the critical XY chain using generalized Fisher-Hartwig theorem. Finally,
 it will be interesting to study the effect of the basis on the behavior of the formation probabilities. In particular, it is interesting to study, for example, 
 the formation probabilities of the Ising model in the $\sigma^x$ basis. Although, thanks to  numerical calculations \cite{NR2015}, we know
 the scaling limit of the configurations in the $\sigma^z$ basis the same problem in the $\sigma^x$ basis is more challenging both analytically and numerically.
 For example, in the Ising model  although we expect that all the $\sigma^x$ spins up configuration leads to a fixed boundary condition  it is not yet clear what will happen if we 
 consider anti-ferromagnetic configuration or more general crystal configurations discussed in \cite{NR2015}.

 \paragraph*{Acknowledgments.} 
 I thank J. Dubail for pointing out to me the reference \cite{StephanDubail}.
 I thank ICTP for hospitality during a period which part of this work was completed.  This work was supported in part by
CNPq.

\begin{appendices}
\section{ Conformal map from a cylinder with two slits to an annulus}

The  conformal map from
an infinite cylinder with two aligned slits (in the periodic direction of the cylinder)  with lengths $s_1$ and $s_2$ and the distance
$l$ to an annulus is as follows\cite{Rajabpour2016}: First,  we introduce a conformal 
map $\tilde{z}(z)$, which takes the system from an infinite cylinder with two slits
to a whole plane with two symmetric aligned slits, with length $\frac{1}{k}-1$ and the distance ~2, as follows:
\begin{eqnarray}\label{conformal map2 details}
\tilde{z}&=&\frac{e^{2i\pi\frac{z}{L}}+a_0}{b_1e^{2i\pi\frac{z}{L}}+b_0},\\
a_0&=&\frac{e^{2i\pi\frac{s_1}{L}}}{N}\Big{(}1-k-2e^{2i\pi\frac{l+s_1}{L}}+(1+k)e^{2i\pi\frac{l}{L}}\Big{)},\nonumber\\
b_1&=&\frac{-1}{N}\Big{(} (1-k)e^{2i\pi\frac{l+s_1}{L}}+2k-(1+k)e^{2i\pi\frac{s_1}{L}}\Big{)},\nonumber\\
b_0&=&\frac{e^{2\pi\frac{s_1}{L}}}{N}  \Big{(}1-k+2ke^{2i\pi\frac{l+s_1}{L}}-(1+k) e^{2i\pi\frac{l}{L}}\Big{)},\nonumber\\
N&=&-2-e^{2i\pi\frac{l+s_1}{L}}(-1+k)+e^{2i\pi\frac{s_1}{L}}(1+k),\nonumber
\end{eqnarray}
with the $k$ given by
\begin{eqnarray}
 k&=&1+2\frac{\sin[\frac{\pi s_1}{L}]\sin[\frac{\pi s_2}{L}]-\sqrt{\sin[\frac{\pi s_1}{L}]\sin[\frac{\pi s_2}{L}]\sin[\frac{\pi (s_1+l)}{L}]
\sin[\frac{\pi (s_2+l)}{L}]}}{\sin[\frac{\pi l}{L}]\sin[\frac{\pi (l+s_1+s_2)}{L}]}.
\end{eqnarray}
The conformal map from the plane with two symmetric slits  to an annulus with the inner 
and outer radiuses $r=e^{-h}$ and $r=1$  has the following form:

\begin{eqnarray}\label{conformal map1}
w(z)&=&e^{-\frac{h}{2}}e^{h \frac{\mbox{sn}^{-1}(\tilde{z},k^2)}{2\mathcal{K}(k^2)}},\\
h&=&2\pi\frac{\mathcal{K}(k^2)}{\mathcal{K}(1-k^2)},
\end{eqnarray}
where $\mathcal{K}$ and $\mbox{sn}^{-1}$ are
the elliptic  and inverse Jacobi functions \footnote{Note that in all of the formulas we adopt
the Mathematica convention for all the elliptic functions.} respectively

\section{Boundary CFT of the critical Ising chain} 
In this appendix we list some necessary formulas for the boundary conformal field theory of the critical Ising model.
There are two different conformal boundary conditions for the critical Ising model, free and fixed boundary conditions \cite{Cardy1989}.
 These two boundary conditions can produce four different
partition functions: 1) fixed with spins in the same direction on both boundaries  "Fi1-Fi1"
2) fixed with spins in the opposite direction "Fi1-Fi2" 3) free on one boundary and fixed on the other one "Fr-Fi" and 4) free on both boundaries "Fr-Fr".
The corresponding partition functions on the cylinder with the length $h$ and the circumference $2\pi$ can be written 
with respect to characters as follows
\begin{eqnarray}\label{partition functions Ising}
Z_{_{Fi1-Fi1}}=\chi_0(\tau)+\chi_{1/2}(\tau)+\sqrt{2}\chi_{1/16}(\tau),\\
Z_{_{Fi1-Fi2}}=\chi_0(\tau)+\chi_{1/2}(\tau)-\sqrt{2}\chi_{1/16}(\tau),\\
Z_{_{Fr-Fr}}=\chi_0(\tau)+\chi_{1/2}(\tau),\\
Z_{_{Fr-Fi}}=\chi_0(\tau)-\chi_{1/2}(\tau),
\end{eqnarray}
where the characters are defined as follows:
\begin{eqnarray}\label{characters}
\chi_0(\tau)=\frac{1}{2\sqrt{\eta(\tau)}}\Big{(}\sqrt{\Theta_3(\tilde{q}^{1/2})}+\sqrt{\Theta_4(\tilde{q}^{1/2})}\Big{)},\\
\chi_{1/16}(\tau)=\frac{1}{2\sqrt{\eta(\tau)}}\sqrt{\Theta_2(\tilde{q}^{1/2})},\\
\chi_{1/2}(\tau)=\frac{1}{2\sqrt{\eta(\tau)}}\Big{(}\sqrt{\Theta_3(\tilde{q}^{1/2})}-\sqrt{\Theta_4(\tilde{q}^{1/2})}\Big{)}.
\end{eqnarray}
where $\Theta_i$'s are the Jacobi theta functions and  $\tilde{q}=e^{\pi i\tau}$ with $\tau=i\frac{h}{\pi}$ is as before.
$\eta$ is the Dedekind function with the following definition
\begin{eqnarray}\label{Dedekind eta}
\eta(q)=q^{\frac{1}{24}}\prod_{n=1}^{\infty}(1-q^n).
\end{eqnarray}
In this paper we just work with $Z_{_{Fr-Fr}}$ and $Z_{_{Fr-Fi}}$.
\end{appendices}

\end{document}